\documentstyle[aps,preprint,epsfig,tighten]{revtex}
\newcommand{\BE}{\begin{equation}}
\newcommand{\EE}{\end{equation}}
\newcommand{\BEA}{\begin{eqnarray}}
\newcommand{\EEA}{\end{eqnarray}}
\newcommand{\Vp}{{\mathbf{p}}}

\begin{document}

\draft

\title{
Intermittent, Domain-Structured Coherence in the Pion Interferometry
of Relativistic Nuclear Collisions
}

\author{Hiroki Nakamura$^{(1)}$ and Ryoichi Seki$^{(2,3)}$}

\address{${}^{(1)}$ Department of Physics,
Waseda University, Tokyo 169-8555, Japan \\
${}^{(2)}$ Department of Physics, California State University, 
Northridge, CA 91330 \\
${}^{(3)}$ W.~K.~Kellogg Radiation Laboratory, Caltech 106-38,
Pasadena, CA 91125}
\date{September 27, 1999}
\maketitle

\begin{abstract}
We investigate a domain-structured source in the pion interferometry 
of relativistic nuclear collisions.  The source emits coherent pions   
intermittently with the background of chaotic pions.
The coherent pions examined are either of a general nature 
or of disoriented chiral condensate.
Two- and three-pion correlations for the source are shown to agree 
well with the recent NA44 experimental data.
\end{abstract}

\bigskip

Two-pion interferometry has been used in relativistic nuclear collisions 
to extract information regarding the size and shape of
the pion-emitting source formed during the collision\cite{rev}.
Pion interferometry based on the Hanbury-Brown-Twiss (HBT) effect is
not limited to that of two pions, but can also be of multi-pions, as in the 
case of three-pion interferometry.  Though multi-pion interferometry is
unavoidably complicated and obviously more difficult experimentally,
it is expected to yield possibly new information that two-pion
interferometry cannot provide\cite{hz,ns1}.

Recently, the NA44 collaboration at CERN has reported a puzzling result 
from the first three-pion interferometry experiment\cite{na44}.
Both the two- and the three-pion correlations are weak, with the three-pion 
correlation nearly vanishing.  The strength of the two- and three-pion 
correlations is measured in terms of the chaoticity $\lambda$ and the  
weight factor $\omega$\cite{hz,ns1,ns2}, the NA44 experiment yielding  
$\lambda = 0.4-0.5$ and $\omega = 0.20 \pm 0.02 \pm 0.19$.  

When the source is completely chaotic, $\lambda$ and $\omega$ are unity, 
while $\lambda$ and $\omega$ vanish for a fully coherent source\cite{note}.  
Over the years, it has been known\cite{gkw} that a source consisting of 
two components, one coherent and the other chaotic, 
a so-called partially coherent source, 
gives $\lambda < 1$.  Such a source would not yield, however,
the NA44 values of $\lambda$ and $\omega$ simultaneously.  In Fig. 1, 
we illustrate a comparison of the partially coherent model\cite{hz} 
with the data.  While $\lambda =$ 0.4 -- 0.5 corresponds to the source
being about 70 \% coherent, the weight factor of 0.20 $\pm$ 0.19 implies
that it is nearly 100\% coherent.  

In this letter we examine a more complicated, believed to be more realistic, 
structure of the pion source that intermittently emits coherent pions from 
several regions with the background of chaotic pion emission. 
We also examine the consequences in pion interferometry when such coherent 
pions are from a domain structure of disoriented chiral condensate (DCC).
We find that both satisfy the NA44 data.

For the pions emitted from several coherent regions with a chaotic
background, we write the source current consisting of multi-components 
as ~\cite{gkw,ns2}
\BE
J^i(x) = \sum_{n=1}^Nj^i(x-X_n)e^{-i\theta_n} + J^i_{cha}(x),
\EE
where $j (x)$ and $J_{cha}(x)$ are coherent and 
chaotic source currents, 
respectively.  The superscript $i$ stands for the charge state of the pions,
such as $+$, $-$, and $0$.
The $n$-th coherent domain is located at $X_n$, and is 
distributed according to $\rho(X_n)$ (appearing below), 
normalized as $\int\rho(x)d^4x =1$.
$\theta_n$ is a random phase, uniformly distributed over the range 
$[ 0, 2\pi)$.  In order to describe an intermittent pion emission from a 
domain-structured source, {\it the coherent sources are taken to obey 
the Poisson distribution}
\BE
{\gamma}_N (\alpha) =\frac{\alpha^N}{N!}e^{-\alpha}.
\EE

We apply a generating functional method for these $c$-number source 
currents\cite{apw}, defined (slightly differently from {}\cite{apw}) as
\BE
G[z_i^*(\Vp),z_i(\Vp)]= \left\langle
 \exp\left\{\sum_{i}\int \frac{d^3{\Vp}}{\sqrt{(2\pi)^3\cdot 2p^0}}
\left(z_i^*(\Vp){J^i}^*(p)+z_i(\Vp)J^i(p)\right)
  \right\} \right\rangle_J,
\label{eq:gf}
\EE
where $p^0$ is on-shell.
By taking functional derivatives of Eq. (3),
one-, two- and three-pion spectra are then constructed as 
\BEA
W^i_1(p_1) &=& \langle | J^i(p_1)|^2 \rangle_J, \\
W^i_2(p_1,p_2) &=& \langle | J^i(p_1)|^2 | J^i(p_2)|^2\rangle_J, \\
W^i_3(p_1,p_2,p_3) &=& 
\langle | J^i(p_1)|^2 | J^i(p_2)|^2| J^i(p_3)|^2\rangle_J, 
\EEA
where $J^i(p)$ is the Fourier transform of $J^i(x)$.
The normalized two- and three-correlation functions are then expressed 
in terms of $W$'s as
\BEA
C^i_2(p_1,p_2)
    &=& \frac{\langle n_{\pi_i} \rangle^2}{\langle n_{\pi_i}(n_{\pi_i}-1)
 \rangle}
        \frac{W^i_2(p_1,p_2)}{W^i_1(p_1)W^i_1(p_2)},\\
\label{eq:c3def}
C^i_3(p_1,p_2,p_3) &=&
  \frac{\langle n_{\pi_i} \rangle^3}{\langle n_{\pi_i}(n_{\pi_i}-1)
(n_{\pi_i}-2) \rangle}
     \frac{W^i_3(p_1,p_2,p_3)}{W^i_1(p_1)W^i_1(p_2)W^i_1(p_3)},
\EEA
where $n_{\pi_i}$ is the number operator of the pions emitted.  

$\langle {\cal O} \rangle_J$
is the statistical average of $\cal O$ about the fluctuation of $J_i(p)$.
\BEA
\label{eq:ave}
\langle{\cal O}\rangle_J
&=&
\sum_{N=1}^\infty {\gamma}_N ({\alpha})
\int \left(
\prod_{i=+,-,0}
{\cal P}_i[J_{cha}^{i*}(p),J_{cha}^i(p)]
{\cal D}J_{cha}^i(p)
{\cal D}J_{cha}^{i*}(p) \right)   
\nonumber \\ &&
\times
\int \left( \prod_{n=1}^N d^4X_N \rho(X_N) \right) 
\int_0^{2\pi}\left(\prod_{n=1}^N \frac{d\theta_n}{2\pi}\right){\cal O}.
\EEA
${\cal P}_i[J_{cha}^{i*}(p),J_{cha}^i(p)]$ is a distribution functional
of $J_{cha}^i(p)$, and assumed to have a Gaussian form, as in Ref.\cite{apw},
so that the higher-order moment of $J_{cha}^i(p)$ is represented
by the second-order moment, for example,
\BEA
\lefteqn{\langle J_{cha}^{i*}(p_1)J_{cha}^{i*}(p_2)J_{cha}^i(p_1)
J_{cha}^i(p_2) \rangle_J =} \nonumber \\
&& \langle J_{cha}^{i*}(p_1)J_{cha}^i(p_1) \rangle_J
 \langle J_{cha}^{i*}(p_2)J_{cha}^i(p_2) \rangle_J
+ \langle J_{cha}^{i*}(p_1)J_{cha}^i(p_2) \rangle_J
 \langle J_{cha}^{i*}(p_2)J_{cha}^i(p_1) \rangle_J.
\EEA
The chaoticity and the weight factor are defined as
\BEA
\lambda_i &=& C^i_2(p,p) -1,\\
\omega_i  &=& \frac{C^i_3(p,p,p)-3\lambda_i-1}{2\sqrt{\lambda_i^3}}.
\EEA

Usually, $\lambda_i$ and $\omega_i$ are independent of the charge state,
$i$, because of charge symmetry.  We will not show $i$ explicitly 
until we discuss the case of DCC.
After some algebra~\cite{ns2}, we obtain the chaoticity 
for the current, Eq. (1), 
\BE
\lambda = \frac{\alpha}{\alpha+(1-\epsilon)^2},
\EE
where $\alpha$ ($=\langle N \rangle$)
 is the mean number of the coherent sources,
and  $\epsilon$ is the ratio of the number of pions emitted from the 
chaotic source and the total number of pions.
When $\alpha \to\infty$ or $\epsilon \to 1$, 
Eq. (8) gives $\lambda \rightarrow 1$.  This condition corresponds 
to a totally chaotic source, in accordance with the description 
of a chaotic source given in \cite{gkw}, 
an infinite number of randomly distributed coherent sources.
The weight factor is obtained, again after some algebra~\cite{ns2}, as
\BE
\omega=
\frac{2\alpha^2+8\alpha(1-\epsilon)^2+3(1-\epsilon)^3(1-2\epsilon)}
{2\left(\alpha^2+3\alpha(1-\epsilon)^2+(1-\epsilon)^3\right)}
\sqrt{\frac{\alpha+(1-\epsilon)^2}{\alpha}}.
\EE
Note that $\omega \to 1$ as $\alpha\to\infty$ or $\epsilon \to 1$.

Figure \ref{fig:MC} illustrates $\lambda$ and $\omega$ when 
the parameters $\alpha$ and $\epsilon$ are varied.  The best fit
to the NA44 data\cite{na44}, $\lambda = 0.45$ and $\omega=0.20$,
corresponds to $\alpha=0.13$ and $\epsilon=0.60$.
This implies that one or two out of ten events contain coherent sources, 
the others being from totally chaotic sources, but about 40\% of the emitted
pions must come from the coherent sources.  That is, very large coherent
sources are sometimes created.  Note that each of the coherent sources
produces, on average, about five times more pions than the chaotic source.

We expect that the coherent sources are generated in association with
a phase transition of quantum chromodynamics.  Depending on specific 
dynamics involved in the generation, the coherent sources would have 
more complicated features and the preceding treatment would require some 
modification.  In the following, as an example, we take a possible phenomenon 
of disoriented chiral condensate (DCC)\cite{Bjorken,rajagopal,gavin}.

The source current is now modified as~\cite{ns3} 
\BE
J^i(p) = \sum_{n=1}^N j^i(p) e^{ip\cdot X_n -i\theta_n}n^i_{n}+J^i_g(p),
\EE
so as to incorporate the proposed consequence of DCC, 
the chiral order-parameter directed in the isospin space,
differently from that in the true vacuum. 
Here, ${\mathbf{n}}_{n}$ is a unit-vector in the isospin space,
describing the direction of condensate of the $n$-th domain.
The statistical average $\langle {\cal O} \rangle_J$ of Eq. (\ref{eq:ave}) 
is also modified to include 
\BE
\int \left(\prod_{n=1}^N
\frac{d^3{\mathbf{n}}_{n}}{4\pi}
\delta(|{\mathbf{n}}_{n}|-1) \right) .
\EE

After lengthy algebra~\cite{ns3}, we obtain the chaoticity and weight factor 
for the positive pions as
\BEA
\lambda_{+} &=& \frac{\alpha}{\alpha+\frac{6}{5}(1-\epsilon)^2}, \\
\omega_{+} &=& \frac{\alpha^2+\frac{6}{5}\alpha(1-\epsilon)^2
    +\frac{27}{175}(1-\epsilon)^3(13-28\epsilon)}
    {\alpha^2+\frac{18}{5}\alpha(1-\epsilon)^2+\frac{54}{35}(1-\epsilon)^3}
 \sqrt{\frac{\alpha+\frac{6}{5}(1-\epsilon)^2}{\alpha}}.
\EEA
For the neutral pions, we obtain      
\BEA
\lambda_{0} &=& \frac{\alpha}{\alpha+\frac{9}{5}(1-\epsilon)^2}, \\
\omega_{0} &=& \frac{1}{2}
  \frac{2\alpha^2+\frac{18}{5}\alpha(1-\epsilon)^2
         +\frac{81}{175}(1-\epsilon)^3(17-42\epsilon)}
      {\alpha^2+\frac{27}{5}\alpha(1-\epsilon)^2
        +\frac{27}{7}(1-\epsilon)^3}
  \sqrt{\frac{\alpha+\frac{9}{5}(1-\epsilon)^3}{\alpha}}.
\EEA

Figure \ref{fig:MDCC} shows that the multiple DCC domain model 
successfully yields the chaoticity and weight factor in agreement with the 
data.  The parameter values for the best fit are found to be 
$\alpha (= \langle N \rangle) = 0.18$ and $\epsilon=0.57$.
This implies that the mean number of the DCC domains
is 0.18 and that the ratio of the chaotic pion number and the total pion 
number is 0.57.  For these parameter values, 
the probability distribution of the ratio of
the neutral pion number and the total pion number, $f$, is 
\BEA
\label{eq:disae}
 P_{\alpha,\epsilon}(f) &=& \sum_{N=0}^\infty \frac{\alpha^N}{N!}e^{-\alpha}
  \int \delta\left( f - \frac{\sum_{n}^N3f_n(1-\epsilon)+\epsilon\alpha}
   {3N(1-\epsilon)+3\epsilon\alpha}\right)
   \prod_{n=1}^N\frac{df_n}{2\sqrt{f_n}} \nonumber \\
                        &\approx& 0.83 P_0(f) + 0.15 P_1(f).
\label{eq:distex}
\EEA
Here, $P_0(f)$ is equal to $\delta(f - 1/3)$ in our simple model, 
while $P_1(f)$ is approximately the inverse square root of $f$. 
Since $P_1(f)$ is suppressed by the factor of 
0.15, $P_{\alpha,\epsilon}(f)$ of Eq. (\ref{eq:distex}) is a distribution 
dominated by a sharp peak with a slow-varying $1/\sqrt{f}$-like background. 
In practice, the sharp peak can be replaced by a smoother function 
such as a binomial distribution peaking at $f=1/3$.

We have thus shown that the NA44 data can be explained by the model of 
multiple DCC with a chaotic pion source.  We have not demonstrated, 
however, that the NA44 data prove the appearance of DCC.  Unfortunately, 
in order to prove this solely by interferometry, we require interferometry 
of the neutral pions. 
For the above parameter values, the chaoticity and weight factor for the 
neutral pions are $\lambda_{0}=0.35$ and $\omega_{0} = -0.12$, respectively.
If interferometry of the neutral pions could be carried out, these 
values should signal the observation of DCC.  

Note that the pion 
interferometry of differently charged pions does not serve for identifying 
DCC because the chaotic part exhibits no HBT effect and the DCC part is
taken to be coherent.

The preceding direct fit to the NA44 data may be, however, associated with
a substantial systematic error.  In interferometry experiments, the events
are often selected by ``minimum bias,'' so as to improve statistics.  
The selection results in combining the events of different multiplicities,
and for more detailed analyses a refinement is needed, such as 
taking account of event-by-event fluctuations. 

We should comment on other possible reasons for the small weight factor
in the experiment.  First, as noted in \cite{ck}, the contamination 
resulting from other particles being regarded as pions
could make the weight factor smaller.  According to our estimate, 
however, the contaminant ratio would have to be as large as about 30\%, 
to obtain an agreement with the data.  We would not expect 
that the accuracy of particle identification be this bad.  
Second, the long-lived resonances could cause the chaoticity
to appear smaller\cite{gp}.  This could happen, but the weight factor 
would not be affected by the decay process of resonances 
since the process is chaotic.  Thus, if this should occur, 
the chaoticity would be smaller than expected, but 
the weight factor would remain as the expected value, contradicting 
the NA44 data.  Third, some complicated final-state interactions 
could induce the apparent result.  While we have not investigated 
all final-state interactions, we note that effects of the major 
final-state interactions were removed in the extraction of 
the NA44 data\cite{na44}.   It is hard for us to expect that some other 
effects might bring about the puzzling data.

In summary,
we introduce a domain-structured coherent source with the background 
of a chaotic source. The pions are emitted intermittently, 
as the number of the domains is different in each event.
We investigate the chaoticity and the weight factor, 
as measures of two- and three-pion correlations from such sources.
While these quantities do not agree with the data in the simple 
partially coherent source model, they do agree in the newly introduced, 
domain-structured coherent-source model. 
Furthermore, when the source is associated with DCC, 
the quantities also agree with the NA44 data.

We acknowledge informative and stimulating discussions with T. Humanic,
especially regarding the NA44 experiment.
This research is supported
by the U.S.~Department of Energy under grant DE-FG03-87ER40347 at CSUN,
and the U.S.~National Science
Foundation under grants PHY88-17296 and PHY90-13248 at Caltech.

\begin{figure}
\begin{center}
\epsfig{file=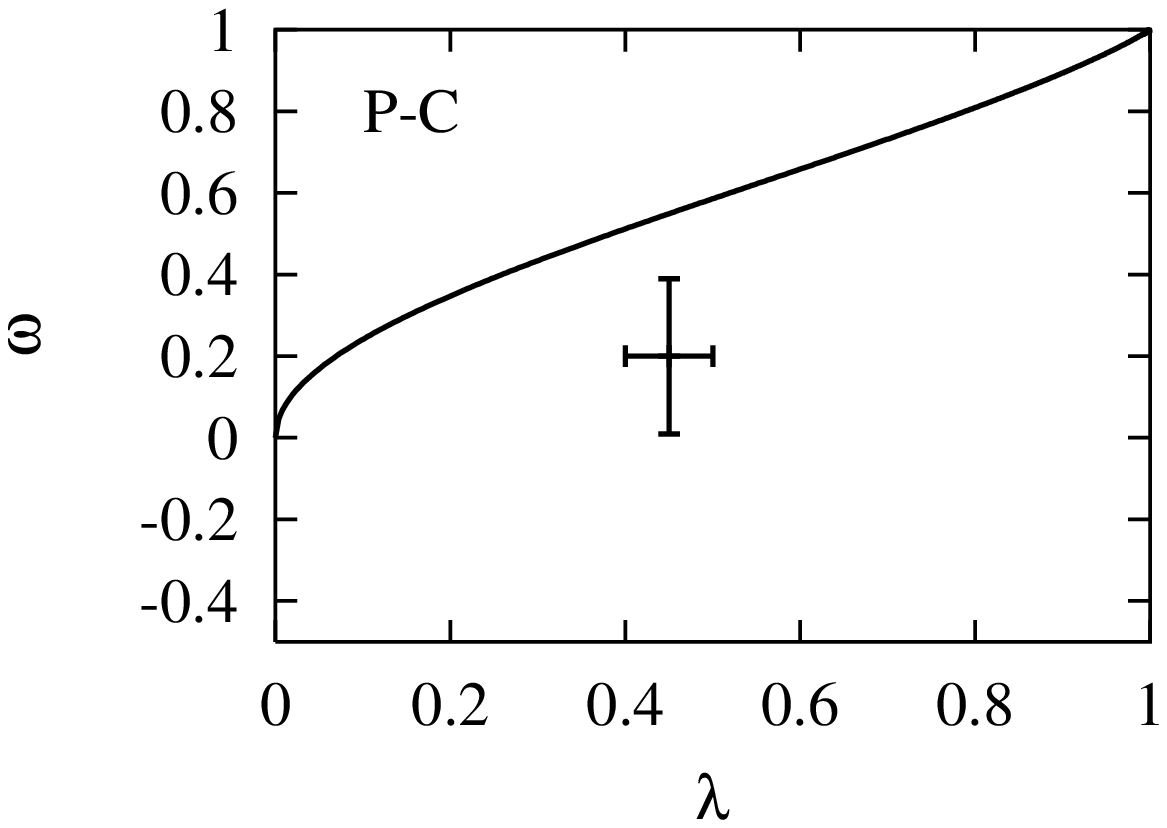}
\caption{Weight factor, $\omega$, as a function of chaoticity, $\lambda$, 
for the model of a partially coherent source (P-C) [2] 
varying the relative coherency.  The data is from the NA44 experiment [4].
}
\label{fig:PC}
\end{center}
\end{figure}

\begin{figure}
\begin{center}
\epsfig{file=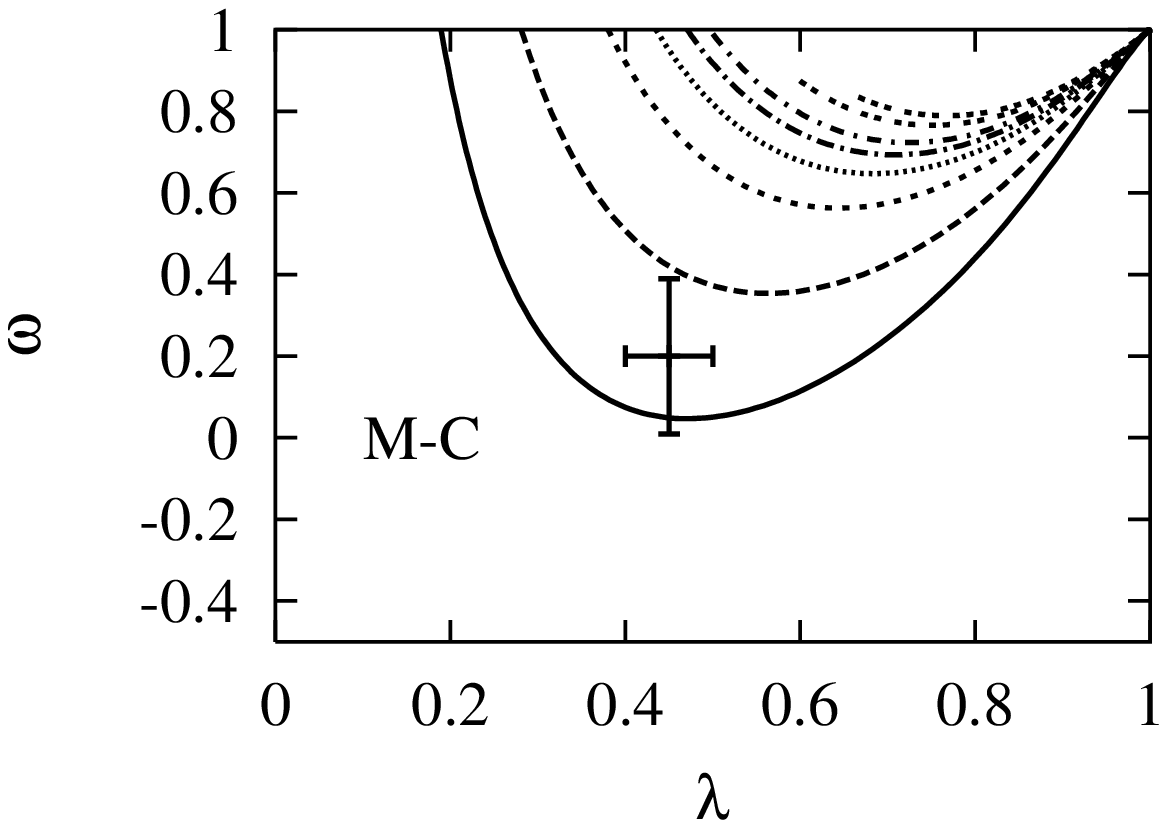}
\caption{Weight factor, $\omega$, as a function of chaoticity, $\lambda$, 
for the positive pions, for the model of multiple coherent domains with one 
chaotic source (M-C), varying 
$\epsilon$ from 0 to 1. 
The lines from down to up correspond to the mean number of domains,
$\langle N \rangle = 0.1$, 0.2, 0.4, 0.6, 0.8, 1.0, 1.5 and 2.0, respectively.
The data is from the NA44 experiment [4].
}
\label{fig:MC}
\end{center}
\end{figure}

\begin{figure}
\begin{center}
\epsfig{file=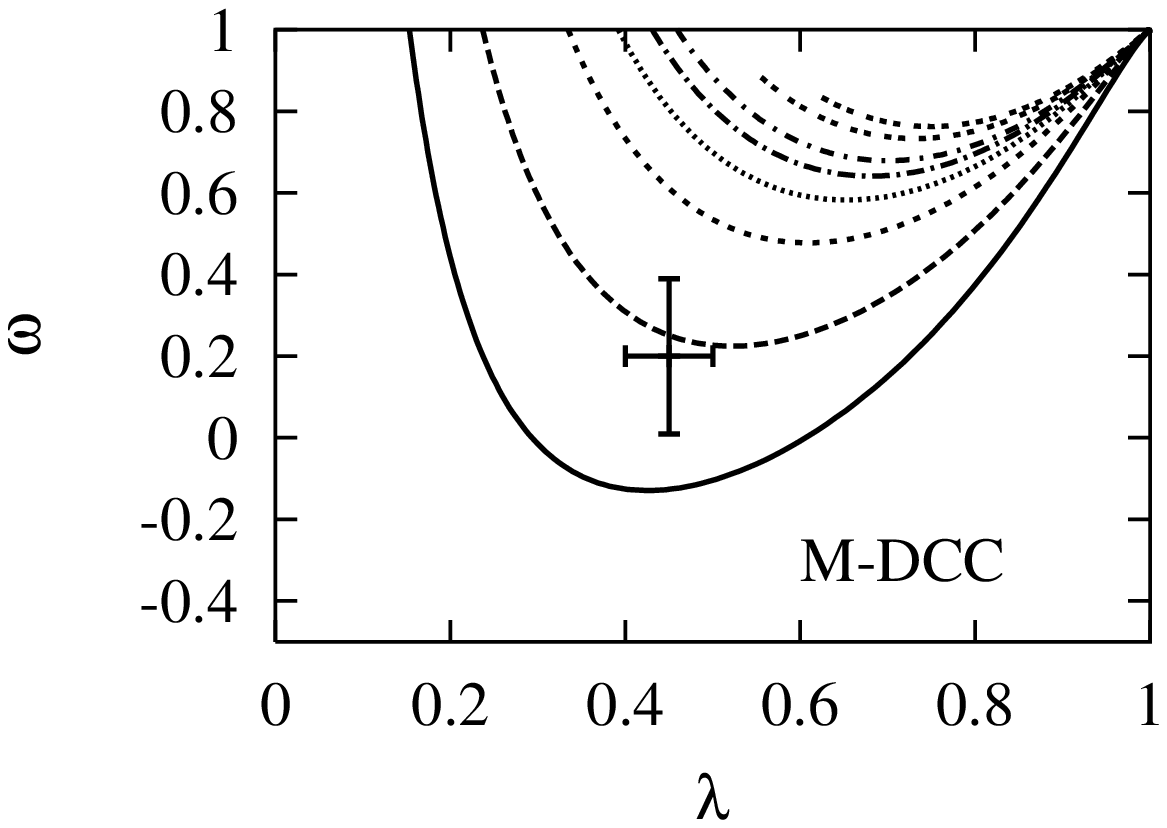}
\caption{Weight factor, $\omega$, as a function of chaoticity, $\lambda$, 
for the positive pions, for the model of multiple DCC domains with one 
chaotic source (M-DCC), varying 
$\epsilon$ from 0 to 1.
The lines from down to up correspond to the mean number of domains,
$\langle N \rangle = 0.1$, 0.2, 0.4, 0.6, 0.8, 1.0, 1.5 and 2.0, respectively.
The data is from the NA44 experiment [4].
}
\label{fig:MDCC}
\end{center}
\end{figure}


\begin{thebibliography}{99}

\bibitem{rev}
{\it Quark Matter '97, Proceedings of the Thirteenth International
 Conference on Ultra-Relativistic Nucleus--Nucleus Collisions},
Tsukaba, Japan, December 1-5, 1997, edited by T. Hatsuda, Y. Miake,
S. Nagamiya, and K. Yagi [Nucl. Phys. {\bf A638}, 1c (1998)].

\bibitem{hz}
U.~Heinz and Q.~H.~Zhang, Phys.\ Rev.\ C{\bf56}, 426 (1997).

\bibitem{ns1}H.~Nakamura and R.~Seki, to appear in Phys. Rev. C,  
nucl-th/9909043.

\bibitem{na44}H.~B{\o}ggild {\it et al.} (NA44 Collaboration),
Phys. Lett. {\bf B455}, 77 (1999).

\bibitem{ns2}H.~Nakamura and R.~Seki, nucl-th/9909044.

\bibitem{gkw}M.~Gyulassy, S. K.~Kauffmann and L. W.~Wilson,
Phys.\ Rev.\ {\bf C20}, 2267 (1979).

\bibitem{note}Strictly speaking, $\omega$ is indefinite for a source being 
100\% coherent.

\bibitem{apw}I.~V.~Andreev, M.~Pl\"{u}mer, and R.~M.~Weiner,
Int. J. Mod. Phys. {\bf A8}, 4577 (1993).

\bibitem{Bjorken}G.~Amelino-Camelia, J.~D.~Bjorken, and S.~E.~Larsson,
Phys.\ Rev.\ {\bf D56}, 6942 (1997); 
J.~D.~Bjorken, Acta Phys. Polon. {\bf B28}, 2773 (1997).

\bibitem{rajagopal}K.~Rajagopal,
in {\it Quark-Gluon Plasma 2}, edited by R. Hwa, World Scientific,
Singapore (1995), hep-ph/9504310; see also the references therein.

\bibitem{gavin}S.~Gavin, BNL-60637, hep-ph/9407368;
Nucl.\ Phys.\ {\bf A590}, 163c (1995);
S.~Gavin, A.~Gocksch, and R.~D.~Pisarski,
Phys.\ Rev.\ Lett.\ {\bf 72}, 2143 (1994).

\bibitem{ns3}H.~Nakamura and R.~Seki, nucl-th/9909045.

\bibitem{ck}
J.~G.~Cramer and K.~Kadija, Phys.\ Rev.\  C{\bf 53}, 908 (1996).

\bibitem{gp}M.~Gyulassy and S. S.~Padula,
Phys.\ Lett.\ {\bf B217}, 181 (1989). 


\end{thebibliography}
\end{document}